\documentclass[iop]{emulateapj}
\usepackage{mathtools}
\usepackage{multirow}

\submitted{Accepted for publication in The Astrophysical Journal (ApJ), 29 September 2016}

\def\gaia{\textit{Gaia}}
\def\hip{\textit{Hipparcos}}
\def\tyc{\textit{Tycho}}

\def\mode{_{Mo}}
\def\rlim{r_{\rm lim}}
\def\rmode{r\mode}
\def\fobs{f_{\rm obs}}
\def\errVarpi{\sigma_\varpi}

\def\bias{\overline{x}}
\def\rms{\overline{x^2}^{1/2}}
\def\stddevmod{\sigma_{x,Mo}}

\def\fr{f_r}
\def\errDist{\sigma_r}

\def\expp{exponentially decreasing space density}
\def\mw{Milky Way}

\begin{document}
\title{Estimating distances from parallaxes. III.~Distances of two million stars in the \gaia{} DR1 catalogue}


\author{Tri L. Astraatmadja$^{1,2}$ and Coryn A. L. Bailer-Jones$^{2}$}
\affil{$^1$Department of Terrestrial Magnetism, Carnegie Institution for Science, 5241 Broad Branch Road, NW, Washington, DC 20015-1305, USA}
\affil{$^2$Max Planck Institute for Astronomy, K\"{o}nigstuhl 17, 69117, Heidelberg, Germany}

\shorttitle{Estimating distances from parallaxes} 
\shortauthors{Astraatmadja and Bailer-Jones}

\begin{abstract}
We infer distances and their asymmetric uncertainties for two million stars using the parallaxes published in the \gaia{} DR1 (GDR1) catalogue. We do this with two distance priors: A minimalist, isotropic prior assuming an exponentially decreasing space density with increasing distance, and an anisotropic prior derived from the observability of stars in a Milky Way model. We validate our results by comparing our distance estimates for 105 Cepheids which have more precise, independently estimated distances. For this sample we find that the \mw{} prior performs better (the RMS of the scaled residuals is 0.40) than the \expp{} prior (RMS is 0.57), although for distances beyond 2\,kpc the \mw{} prior performs worse, with a bias in the scaled residuals of -0.36 (vs. -0.07 for the \expp{} prior). We do not attempt to include the photometric data in GDR1 due to the lack of reliable colour information. Our distance catalogue is available at \url{http://www.mpia.de/homes/calj/tgas\_distances/main.html} as well as at CDS. This should only be used to give individual distances. Combining data or testing models should be done with the original parallaxes, and attention paid to correlated and systematic uncertainties.
\end{abstract}


\section{Introduction}
The ESA \gaia{} mission \citep{gaiamission} is obtaining highly accurate parallaxes and proper motions of over one billion sources brighter than $G\simeq 20.7$. 
The first data release (\gaia{} DR1), based on early mission data, was released to the community on 14 September 2016 (\citealt{gdr1paper}). The primary astrometric data set in this release lists the positions, parallaxes, and proper motions of 2\,057\,050 stars which are in the \tyc{}-2 \citep{hoeg00} catalogue
(93\,635 of the these are \hip{} \citep{per97,van07} sources).
This data set is called the \tyc{}-\gaia{} astrometric solution (TGAS; \citealt{mic15,gdr1astrometry}). 

The 5-parameter astrometric solutions for TGAS stars were obtained by combining \gaia{} observations with the positions and their uncertainties of the 
\tyc{}-2 stars (with an observation epoch of around J1991) as prior information \citep{gdr1astrometry}.
This was necessary because the observation baseline in the early \gaia{} data was insufficient for a 
\gaia{}-only solution.
The resulting solutions have median parallax uncertainties of $\sim$0.3\,mas, with an additional systematic uncertainty of about $\sim$0.3\,mas \citep{gdr1paper, gdr1astrometry}. 

Using the TGAS parallaxes $\varpi$ and uncertainties $\errVarpi{}$, we here infer the distances to all TGAS stars along with (asymmetric) distance uncertainties (as Bayesian credible intervals). The motivation and methods to estimate distances from parallaxes have been described in our earlier works (\citet{cbj15, ast16}, henceforth Paper~I and Paper~II respectively). We will not repeat the discussion here, except to remind readers that inverting parallaxes to estimate distances is only appropriate in the absence of noise. As parallax measurements have uncertainties---and for many TGAS stars very large uncertainties---distance estimation should always be treated as an inference problem.

\section{Properties of TGAS parallaxes and their measurement uncertainties}
\begin{figure*}[t]
$\vcenter{\hbox{\includegraphics[width=0.7\hsize]{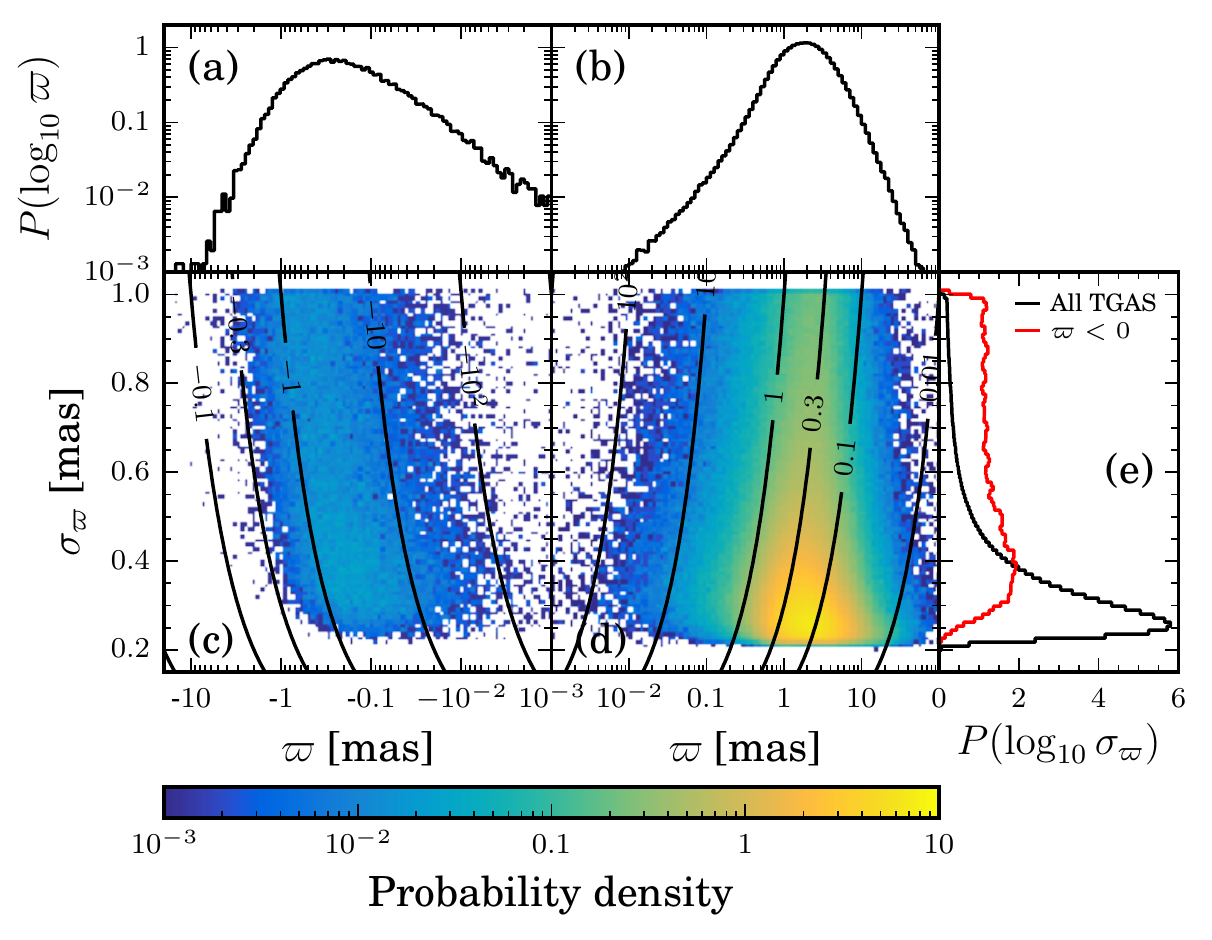}}}$
$\vcenter{\hbox{\includegraphics[width=0.3\hsize]{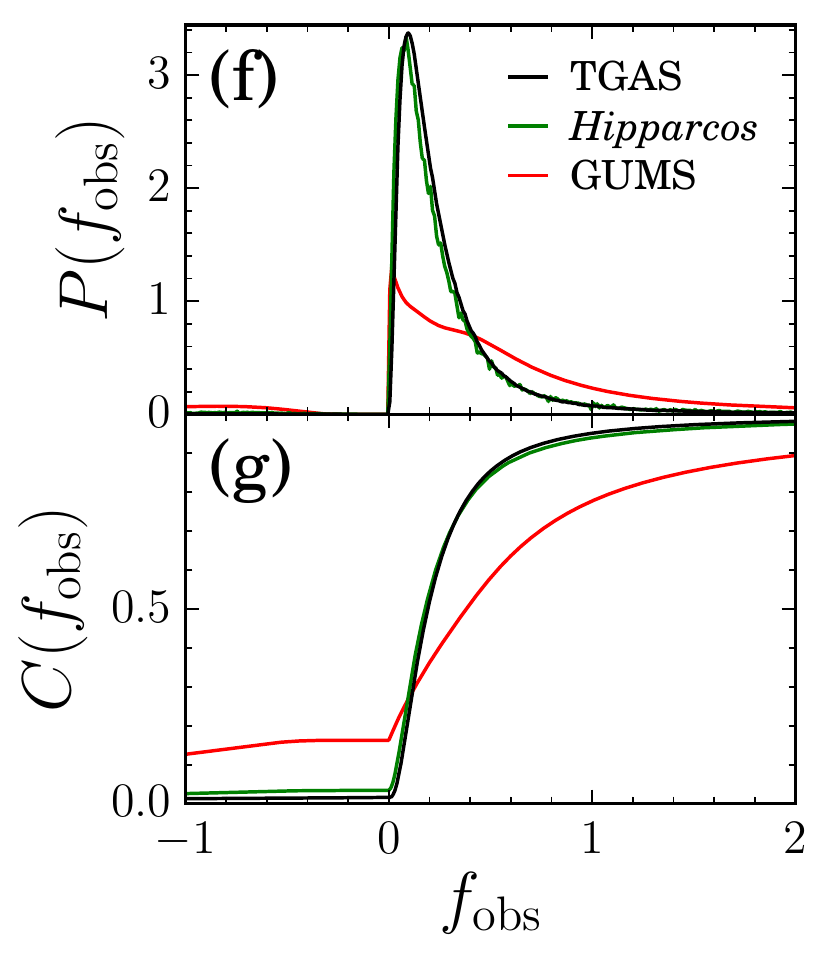}}}$
\caption{The TGAS parallax data. Panels (a) and (b) show the histograms of the TGAS parallaxes, $\varpi$, for negative and positive parallaxes respectively. 
Panels (c) and (d) show the distribution of TGAS parallax uncertainties, $\errVarpi$, as a function of $\varpi$, on a density scale, again for negative and positive parallaxes.
The contour lines show the loci of constant $\fobs = \errVarpi/\varpi$ as indicated by the labels. 
Panel (e) shows the histogram of $\errVarpi$ for all stars (black histogram) as well as the subset which have negative parallaxes (red histogram). The histogram for only positive parallaxes is almost exactly the same as those for all stars and thus is not shown. 
The vertical axes in panels (a) and (b) is logarithmic whereas it is linear in panels (c)--(e).
Panel (f) shows the probability density of the observed fractional parallax uncertainty, $\fobs = \errVarpi/\varpi$, for TGAS stars (black line), compared with \hip{} (green line) and GUMS (red line) stars. Panel (g) shows the corresponding cumulative distributions. Note that the Panels (f)--(g) only cover a subrange of all possible $\fobs$. 
}
\label{fig:varpi_errVarpiDist}
\end{figure*}
\begin{figure*}[t]
\includegraphics[width=\hsize]{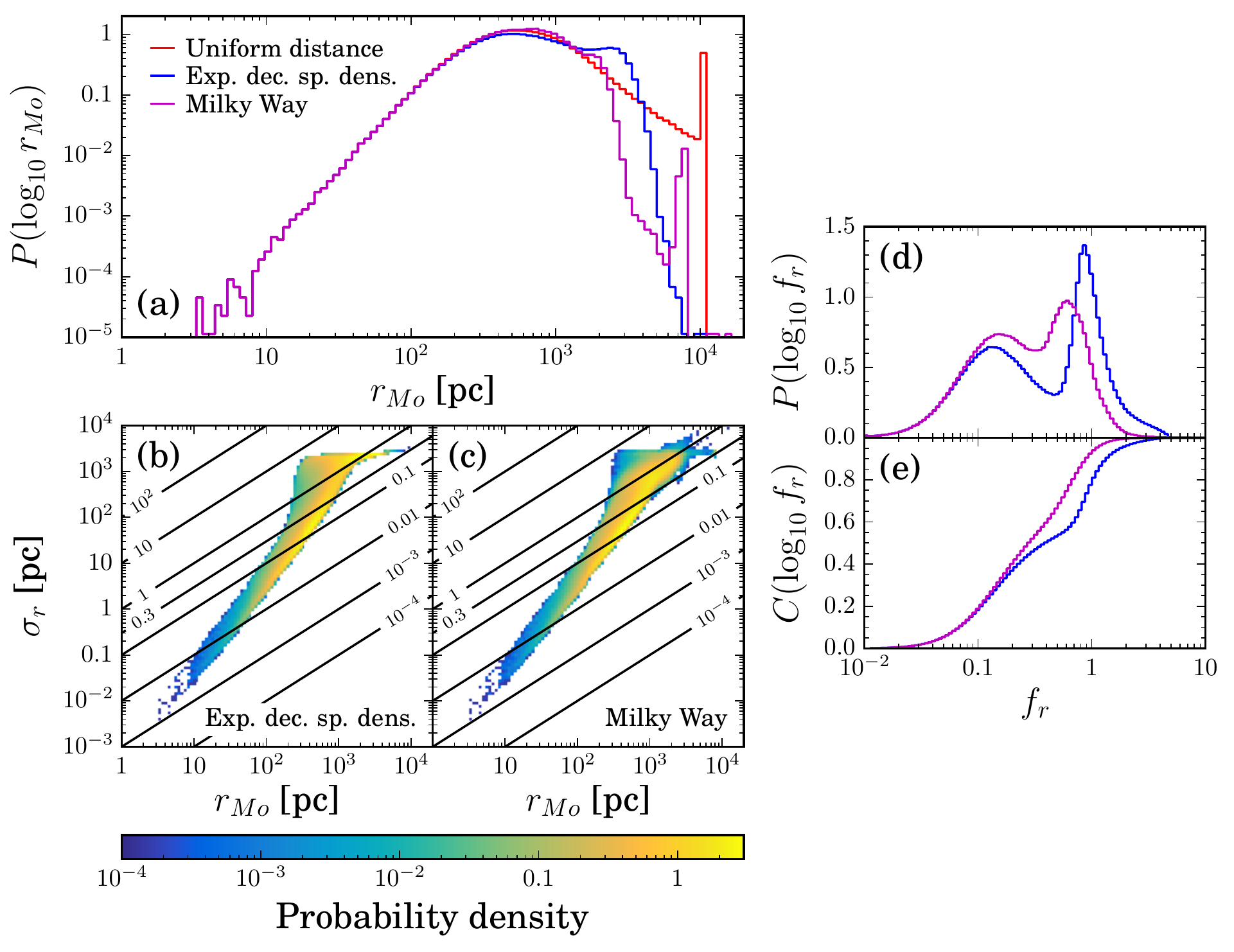}
\caption{Distance estimates for TGAS stars. Panel~(a) shows the distribution of the estimated distances $\rmode{}$ derived from the mode of the three posteriors indicated by the legend. 
Panels~(b) and(c) show the distribution of the distance uncertainties, $\errDist$, as a function of $\rmode$, 
on a density scale, for the \expp{} and the \mw{} priors respectively. The diagonal lines show the loci of constant $\fr = \errDist/\rmode$ as indicated by the labels. Panel~(d) shows the probability distribution functions of the fractional distance uncertainty $\fr = \errDist/\rmode$ for the
\expp{} prior (blue) and the \mw{} prior (magenta). Panel~(e) shows the corresponding cumulative distributions.}
\label{fig:rMoDist}
\end{figure*}
\begin{figure*}[t]
\includegraphics[width=\hsize]{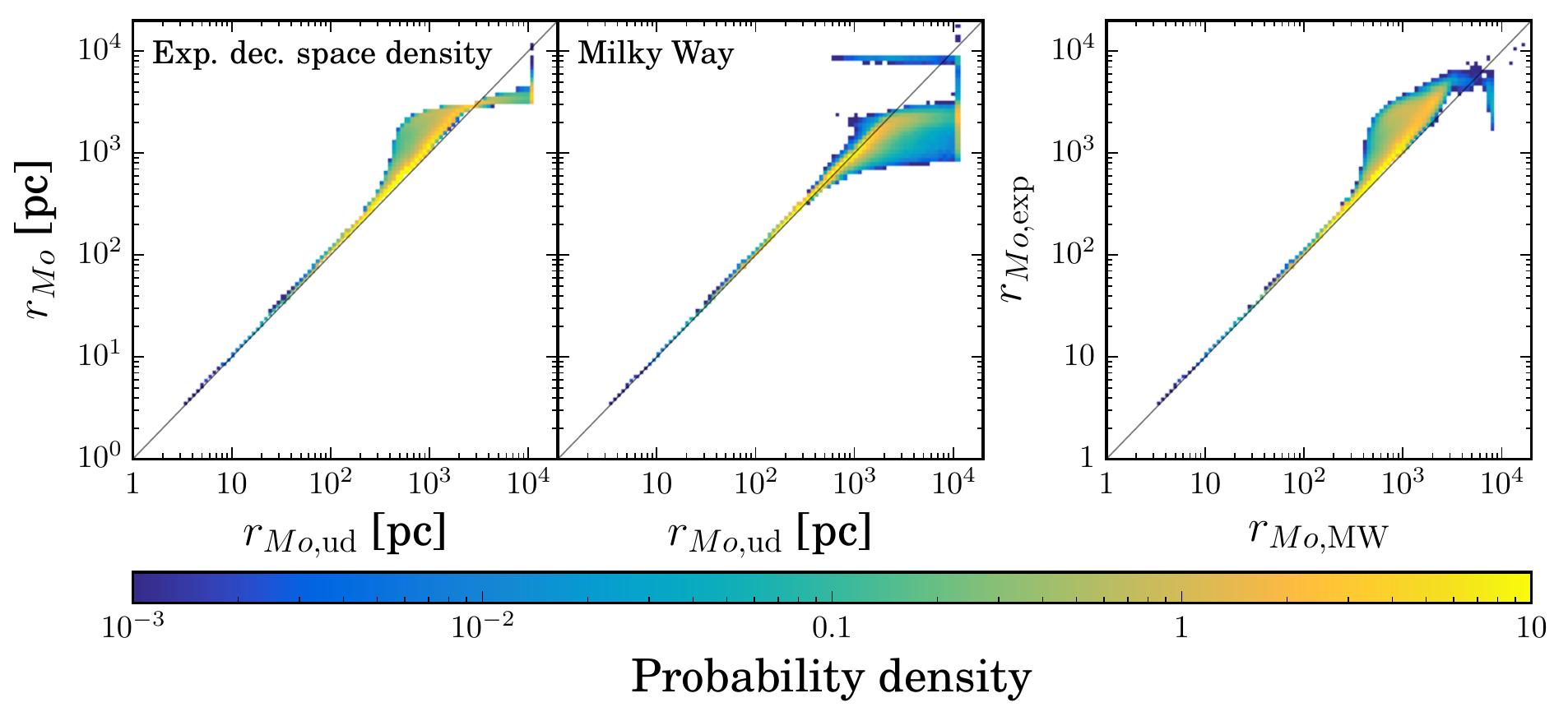}
\caption{Distance estimate comparisons. The left two panels compare our distance estimates from the 
\expp{} and \mw{} priors (vertical axes) with those obtained from the naive uniform distance prior (horizontal axis).
The right panel compares the estimates from our two pain priors, \expp{} prior (vertical axis) and \mw{} prior (horizontal axis).}
\label{fig:rMoComp}
\end{figure*}
Panels (a)--(e) of Fig.~\ref{fig:varpi_errVarpiDist} show the distribution of $\errVarpi$ as a function of $\varpi$, as well as histograms of $\varpi$ and $\errVarpi$. The distribution in $\errVarpi$ covers a narrow range between 0.2 mas and 1 mas (cf. Fig.~13 of Paper~II which shows the same plot for GUMS data\footnote{GUMS, the \gaia{} Universe Model Snapshot \citep{rob12}, is a mock catalogue which simulates the expected content of the final \gaia{} catalogue}), which reflects the preliminary nature of GDR1. The upper limit of 1 mas is due to the imposed $\errVarpi=1$\,mas cutoff to reject unreliable astrometric solutions, while the lower limit is due to the $\sim$0.2\,mas noise floor which is dominated by the satellite attitude and calibration uncertainties \citep{gdr1astrometry}. 
Future data releases will be much more precise \citep{gdr1paper}.

In Panels (f)--(g) of Fig.~\ref{fig:varpi_errVarpiDist} we show the distribution of the fractional parallax uncertainties $\fobs=\errVarpi/\varpi$ of TGAS stars, compared with $\hip$ and GUMS stars. We see here that interestingly the combination of TGAS $\varpi$ and $\errVarpi$ produces a distribution of $\fobs$ that is similar to those of $\hip$ stars.

\section{Method, priors, and data products}

The inferred distances of stars depend not only on the observed parallaxes and their uncertainties, but also on the prior. In this paper we infer distances using two priors: a minimalist, isotropic \expp{} prior and a more complex, anisotropic \mw{} prior. The properties of the \expp{} have been discussed in Paper~I, and in Paper~II we have seen that for an end-of-mission \gaia{}-like catalogue, the optimum scale length $L$ is 1.35\,kpc. We use this value to derive distances here, even though it is optimised for the end-of-mission catalogue, so TGAS stars may have a different true distance distribution.

Although not analysed here, in our catalogue we also provide distances using the \expp{} prior using $L=0.11$\,kpc. This value is found by fitting the prior with the true distance distribution of GUMS stars with $V<11$ (this is the $V$-band magnitude at which $\tyc$-2 is 99\% complete).

The derivation and parameters of the \mw{} prior have been discussed in Paper~II, and illustrations of the resulting posterior for several parallaxes $\varpi$ and uncertainties $\errVarpi{}$ can be seen in Figs.~6--7 of Paper~II. Here we retain the parameters of the \mw{} model as well as the \cite{dri03} extinction map, with the exception of the limiting magnitude $m_{G,{\rm lim}}$ (Eq.~6 in Paper~II), used to calculate the faint end of the luminosity function. In this paper we use $m_{G,{\rm lim}} = 12.998$, which is the 99.9\% percentile of the magnitude distribution of all TGAS stars.

For every single star we compute the posterior PDF over distance.
The distance estimate we report here is the mode of the posterior, $\rmode$. We do not report the median distance because, as we have seen in Paper~II, it is a worse estimator for the priors used here.

In addition to the median we report in our catalogue the 5\% and 95\% quantiles of the posterior, $r_5$ and $r_{95}$. Note that many of the posteriors are asymmetric about the mode (and mean and median).
The difference between these gives a 90\% credible interval, which we then divide by a factor $2s$ to produce
\begin{equation}
\errDist = \frac{r_{95}-r_{5}}{2s},
\end{equation}
where $s=1.645$ is ratio of the 90\% to 68.3\% credible interval \textit{in a Gaussian} distribution.
Thus $\errDist$ is a simplified (symmetric) uncertainty in our distance estimate which is equivalent, in some sense, to a 1$\sigma$ Gaussian uncertainty.

We use neither apparent magnitudes nor colours to help infer the distance, even though we have shown in Paper~II that this significantly improves the distance estimation in many cases. This is because GDR1 does not contain colour information. We chose not to use the \tyc{} photometric data on account of its low precision (median photometric uncertainties in $B_T$ and $V_T$ are respectively 136 and 96 mmag).

In the analyses that follow we have not included in our inference the $\sim$0.3\,mas systematic uncertainties reported for the TGAS parallaxes. This is partly because we know this to be a very rough estimate of the systematics, and is possibly overestimated. But we do provide a second catalogue on the web site mentioned which includes this systematic error. It is included by adding it in quadrature with the random parallax error and then repeating the inference. In general this affects both the mode of the posterior (the distance estimate) and its quantiles (the uncertainty).

\section{Distance estimation results}
\begin{table}[t]
\caption{Statistical summary of the distance estimation of 2 million sources in the primary data set of GDR1. Columns with headings 10\%, 50\%, and 90\% give the lower decile, median, and upper decile of the fractional uncertainty $\fr$ for all 2\,057\,050 sources in the primary data set as well as a subset of 93\,635 sources in common with \hip{}.}
\label{tab:statsum}
\centering
\begin{tabular}{lrrrrrr}
\tableline\tableline
\multicolumn{1}{c}{\multirow{2}{*}{Data set}} &
\multicolumn{3}{c}{TGAS} &
\multicolumn{3}{c}{\hip{} subset}\\
&
\multicolumn{1}{c}{10\%} &
\multicolumn{1}{c}{50\%} &
\multicolumn{1}{c}{90\%} &
\multicolumn{1}{c}{10\%} &
\multicolumn{1}{c}{50\%} &
\multicolumn{1}{c}{90\%}\\ 
\tableline
\multicolumn{7}{c}{Exponentially decreasing space density}\\
All stars         & 0.067 & 0.378 & 1.315 & 0.021 & 0.078 & 0.656\\
$\rmode<200$\,pc  & 0.023 & 0.045 & 0.095 & 0.021 & 0.077 & 0.365\\
\tableline
\multicolumn{7}{c}{\mw{}}\\
All stars         & 0.066 & 0.273 & 0.874 & 0.021 & 0.077 & 0.365\\
$\rmode<200$\,pc  & 0.023 & 0.046 & 0.096 & 0.013 & 0.035 & 0.069\\
\tableline
\end{tabular}
\end{table}
The results of the distance estimation are shown in Fig.~\ref{fig:rMoDist} and the statistics of the uncertainties are summarised in Tab.~\ref{tab:statsum}.
In Panel (a) of Fig.~\ref{fig:rMoDist} we show the distribution of the estimated distance $\rmode$ derived from the mode of the two posteriors already mentioned.
The red line in that panel is for a third posterior which uses the uniform distance prior (Paper~I), with a large cut-off at $\rlim = 10$\,kpc. This posterior is equivalent to inverting the parallax to get a distance, except for the cases where the parallax is very small or negative, in which case the mode of the posterior is at $\rlim = 10$\,kpc. This is the reason for the peak in the distribution we see in Panel (a). It contains 43\,673 stars, which is 2.1\% of TGAS. For the \expp{} prior, we also see a peak, but at around $\rmode = 2.7$\,kpc (it's not very visible as a peak due to the log scale). This is the mode of that prior $(r=2L)$, and the mode of the posterior is very close to this for stars with large parallax uncertainties. The \mw{} prior also has a mode, but because it is an anisotropic prior, the mode varies with line-of-sight direction. However, the most prominent peak at $\rmode\sim 8$\,kpc can be seen, which corresponds to the prior for stars toward the Galactic centre, and thus for poorly measured stars in this direction.

For distances up to about 200\,pc, the distributions of $\rmode$ for both priors are similar to each other. Looking again at Panel~(d) of Fig.~\ref{fig:varpi_errVarpiDist}, we see that for stars with $\varpi\gtrsim 5$\,mas, most stars have
$\fobs<0.2$. We showed in Paper~II that for stars with positive parallaxes and $\fobs\lesssim 0.2$, the distance estimate is largely independent of the choice of prior. Beyond 200\,pc, however, the $\rmode$ distributions for all priors diverge. For distances of more than 1\,kpc, most stars have $\fobs\gtrsim 0.3$ and the distance estimate becomes much more prior-dependent.

The distribution of the fractional uncertainties in distance $\fr = \errDist/\rmode$ is shown in Panels~(d) and (e) of Fig.~\ref{fig:rMoDist}. For both priors, the combined distribution of $\rmode$ and $\errDist$ is similar for $\fr\lesssim 0.1$. Both distributions peak at about 0.15, but beyond that a second peak corresponding to poorly measured stars can be seen at $\fr\sim 0.8$ and $\fr\sim 0.6$ for the \expp{} prior and the \mw{} prior respectively.

We compare the distances estimated using the two priors with each other, and with distances estimated from the uniform distance prior, in Fig.~\ref{fig:rMoComp}. We see again that for distances up to $\sim$200\,pc, distances using all priors are similar. For $1/\varpi\gtrsim 200$\,pc, we start to see elongations that correspond to the mode of the respective priors, as discussed above. 

\section{Validation with Cepheid variables}
\begin{figure}[t]
\includegraphics[width=\hsize]{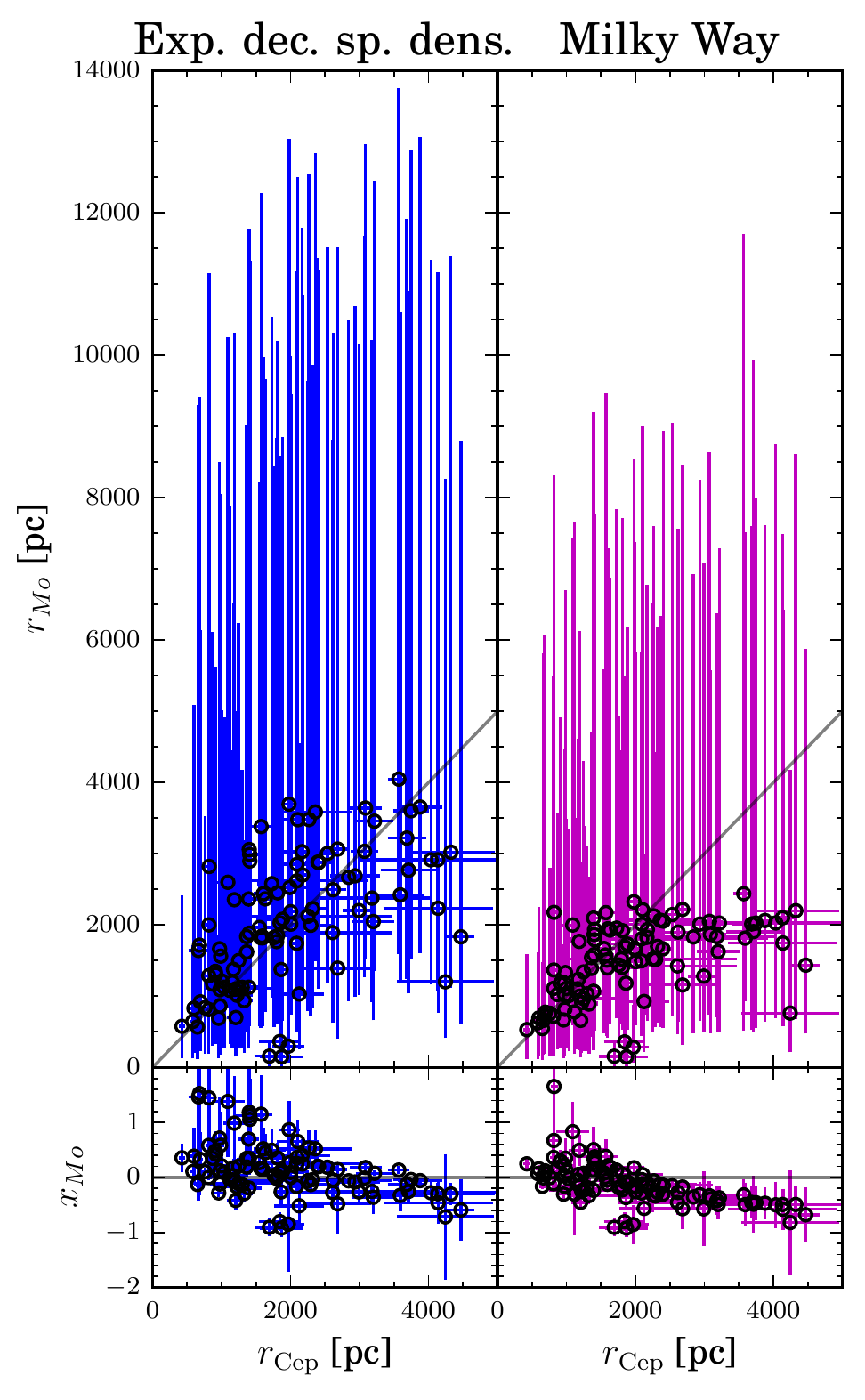}
\caption{A comparison of the distances estimated using the period-luminosity relation of Cepheid stars \citep{gro13} in common with TGAS sources, with the distances of the same stars estimated using the \expp{} (left column) and the \mw{} (right column) prior. The top row shows the distance comparisons. The diagonal lines indicate perfect match between the distances. The bottom row shows the scaled residual $x_{\mode}$ as a function of the Cepheid distance. The horizontal lines indicate zero residuals. The error bars of the estimated distances are the 90\% credible intervals, while for the Cepheids they are the quoted 1$\sigma$ uncertainties multiplied by $s=1.645$ to scale them into the 90\% credible intervals.}
\label{fig:cepDist}
\end{figure}
To see how consistent our estimated distances are with other, more precise, estimates (for distant stars), we compare our estimated distances with the distances of Cepheid variable stars. We took 170 Cepheids from \cite{gro13} and cross-matched them with GDR1 using Simbad. We found 105 Cepheids in common with GDR1. The \cite{gro13} Cepheids have median fractional uncertainties of about $\sim$0.054. Almost all of these Cepheids are \hip{} sources.

Fig.~\ref{fig:cepDist} compares our distances estimates (for both priors) with those of \cite{gro13} for both priors. The bottom row of that figure shows this using the scaled differences
\begin{equation}
x_{Mo} = \frac{\rmode - r_{\rm Cep}}{r_{\rm Cep}} \ .
\end{equation}
The uncertainties in $r_{\rm Cep}$ are taken from \cite{gro13}, where they were computed in a Monte Carlo simulation which takes into account uncertainties in the spectrophotometry, the projection factor, and the phase measurements. 
We multiply these uncertainties by $s=1.645$ to scale them to be 90\% credible intervals, in order to make a fair comparison with our 90\% credible intervals, $r_{95}-r_{5}$. 

To summarize the differences seen in Fig.~\ref{fig:cepDist}, we calculate the bias $\bias$, root mean square (RMS) $\rms$ of the scaled residuals, as well as the standard deviation $\stddevmod$ of the scaled residuals, for all Cepheids, for both priors. 
We also do this separately for near ($r_{\rm Cep}<2$\,kpc) and distant ($r_{\rm Cep}\geq 2$\,kpc) Cepheids.
These results are summarised in Tab.~\ref{tab:statCep}.

Inspecting Fig.~\ref{fig:cepDist} and Tab.~\ref{tab:statCep}, and assuming the \cite{gro13} distances to be ``true'' (for simplicity), we see that overall the \mw{} prior performs better than the \expp{} prior in terms of having a smaller RMS and standard deviation. It is slightly less biased than the \expp{} prior although the bias is in the opposite direction: it tends to underestimate distance. This is due to the assumptions the \mw{} prior makes in the face of poor data, which is that a star is more likely to reside in the disc than further away. Hence this prior becomes mismatched when we only consider the distant Cepheids ($r_{\rm Cep}\geq 2$\,kpc). Distance estimate using the \mw{} prior have a bias of -0.36 for these stars, as is also apparent from Fig.~\ref{fig:cepDist}. For $r_{\rm Cep}\geq 2$\,kpc, when the data are poor, the posterior based on this prior has a mode at around about 2\,kpc, which roughly corresponds to the radial scale length of the thick disk in our \mw{} model. For Cepheids closer than 2\,kpc, however, we see that the \mw{} prior performs well in terms of bias, RMS, and standard deviation. 

The \mw{} prior also gives a more reasonable credible interval than the \expp{} prior, as can be seen in the top row of Fig.~\ref{fig:cepDist}. Most of our TGAS-based distance uncertainties are large, because the Cepheids are distant and have large fractional parallax uncertainties, with median $\fobs$ of about 0.48 (vs.~$\sim$0.2 for all TGAS stars).
Furthermore, the posteriors---and therefore the credible intervals---are highly asymmetric, with a long tail to large distances. This is a natural consequence of the nonlinear transformation from parallax to distance.

The stars used in this validation are intrinsically bright and relatively distant compared to the typical Milky Way stars used to build the \mw{} prior. Our distance estimation is based solely on measured parallaxes; no photometry is involved. 
Thus the \mw{} prior is not well-matched: in the absence of precise parallaxes it tells us that stars are more likely to be in the disc than further away. This explains the poorer behaviour of this prior for distant Cepheids.
The \expp{} performs better in this regime due the scale length $L$ adopted, which puts the mode of the prior at $2L=2.7$\,kpc.
\begin{table}[t]
\caption{The bias $\bias$ as well as the root mean square (RMS) $\rms$ and standard deviation $\stddevmod$ of the scaled residuals of Cepheids stars in the TGAS catalogue.}
\label{tab:statCep} 
\centering
\begin{tabular}{lrrr}
\tableline\tableline
\multicolumn{1}{c}{Prior and sample} &
\multicolumn{1}{c}{$\bias$} &
\multicolumn{1}{c}{$\rms$} &
\multicolumn{1}{c}{$\stddevmod$}\\
\tableline
\multicolumn{4}{l}{Exponentially decreasing space density}\\
\hspace{5mm}All Cepheids & 0.151 & 0.567 & 0.547\\
\hspace{5mm}Cepheids with $r_{\rm Cep}<2$\,kpc & 0.298 & 0.678 & 0.608\\
\hspace{5mm}Cepheids with $r_{\rm Cep}\geq 2$\,kpc & -0.070 & 0.340 & 0.333\\
\multicolumn{4}{l}{\mw{}}\\
\hspace{5mm}All Cepheids & -0.133 & 0.404 & 0.382\\
\hspace{5mm}Cepheids with $r_{\rm Cep}<2$\,kpc & 0.022 & 0.395 & 0.394\\
\hspace{5mm}Cepheids with $r_{\rm Cep}\geq 2$\,kpc & -0.364 & 0.418 & 0.205\\
\tableline
\end{tabular}
\end{table}

\section{Conclusions}
We have inferred the distances of two million stars in the \gaia{} DR1 catalogue using Bayesian inference. The priors used are the \expp{} prior with scale length $L=1.35$\,kpc, and the \mw{} prior with the same parameters as in Paper~II. The median fractional distance uncertainties ($\fr=\sigma/r_{Mo}$) are 0.38 and 0.27 for the \expp{} and the \mw{} prior respectively. If we only consider stars with the estimated distances $\rmode<200$\,pc, the median value of $\fr$ improves at about $\sim$0.04 for both priors. 
This applies to about 193\,000 stars (the exact number is different for both priors) or about 9\% of TGAS.

We validate our distance estimates using more precise distances for Cepheid stars in TGAS taken from \cite{gro13}. We found that for distances closer than 2000\,pc, the \mw{} prior performs better than the \expp{} prior. Beyond 2000\,pc, the \mw{} prior performs worse for this sample (which are intrinsically bright and distant stars) because it assumes that stars are more likely to be closer in the disc than further away. Our \expp{} prior has a longer scale length and thus performs better on this sample when faced with the same poor measurements. But overall the \mw{} prior performs better.

Due to the lack of reliable colours, we do not use these in combination with the parallaxes to estimate distances. Rather than using the \tyc{} magnitudes, significant improvements can be achieved taking spectrophotometric information from other surveys. We choose here just to present astrometric distances.

The distance estimates presented in this paper are useful for \textit{individual} stars. To obtain the mean distance to a group of stars, such as a cluster, one should do a combined inference using the original parallaxes and taking into account the correlated parallax uncertainties for stars observed in a small field. Note, however, that this combination will still not reduce the uncertainty in the mean below the limit presented by the TGAS systematic parallax error. Similarly, if one wishes to compare a model for distances to the TGAS data, this is normally best done by projecting the model-predicted distances into the parallax domain, rather than to use individual estimated distances.

\acknowledgements
This work has made use of data from the European Space Agency (ESA) mission {\it Gaia} (\url{http://www.cosmos.esa.int/gaia}), processed by the {\it Gaia} Data Processing and Analysis Consortium (DPAC, \url{http://www.cosmos.esa.int/web/gaia/dpac/consortium}). Funding for the DPAC has been provided by national institutions, in particular the institutions participating in the {\it Gaia} Multilateral Agreement. We also made use of NASA's Astrophysics Data System; the SIMBAD database, operated at CDS, Strasbourg, France; \texttt{matplotlib}, a Python library for publication quality graphics \citep{Hunter:2007}; and \texttt{TOPCAT}, an interactive graphical viewer and editor for tabular data \citep{2005ASPC..347...29T}.

\bibliographystyle{aasjournal}
\bibliography{bibliography}
\end{document}